\documentstyle[emulateapj,psfig]{article}
\submitted{The Astrophysical Journal Letters, Vol. 528, L13 (2000), in press}
\lefthead{}
\righthead{Imaging a Black Hole}

\begin{document}

\title{Viewing the Shadow of the Black Hole at the Galactic Center}

\author{Heino Falcke\altaffilmark{1}, Fulvio Melia\altaffilmark{2,4}, and Eric 
Agol\altaffilmark{3}}
\altaffiltext{1}{Max-Planck-Institut f\"ur Radioastronomie, Auf dem H\"ugel
69, D-53121, Bonn, Germany}
\altaffiltext{2}{Physics Department and Steward
Observatory, The University of Arizona, Tucson, AZ 85721}
\altaffiltext{3}{Physics and Astronomy Department, Johns Hopkins University,
Baltimore, MD 21218.}
\altaffiltext{4}{Presidential Young Investigator and Sir Thomas Lyle Fellow}
\begin{abstract}
In recent years, the evidence for the existence of an ultra-compact
concentration of dark mass associated with the radio source Sgr A* in
the Galactic Center has become very strong. However, an
unambiguous proof that this object is indeed a black hole is still
lacking.  A defining characteristic of a black hole is the event
horizon.  To a distant observer, the event horizon casts a relatively
large ``shadow'' with an apparent diameter of $\sim10$ gravitational
radii due to bending of light by the black hole, nearly
independent of the black hole spin or orientation.  The predicted
size ($\sim30 \mu$arcseconds) of this shadow for Sgr A*
approaches the resolution of current radio-interferometers.  If the
black hole is maximally spinning and viewed edge-on, then the 
shadow will be offset by $\sim8$ $\mu$arcseconds from the center of mass, 
and will be slightly flattened on one side.  Taking into account
scatter-broadening of the image in the interstellar medium and the
finite achievable telescope resolution, we show that the shadow of Sgr A*
may be observable with very long-baseline interferometry at sub-millimeter 
wavelengths, assuming that the accretion flow is optically thin in this
region of the spectrum. Hence, there exists a realistic
expectation of imaging the event horizon of a black hole within the
next few years.
\end{abstract}


\section{Introduction}
High resolution spectroscopy (especially with the Hubble Space
Telescope) of galactic nuclei has produced an abundance of evidence for
compact dark mass concentrations of up to $10^8 M_\odot/{\rm pc}^3$,
whose nature is strongly suspected to be indicative of supermassive
black holes (\cite{1995ARA&A..33..581K}).  Even better evidence 
exists for the galaxy NGC 4258 and the Milky Way, for which spectroscopic 
and proper motion studies have provided an unprecedented three-dimensional
view of the kinematics of gas and stars around a central point mass,
pointing to dark mass concentrations of $>10^{12} M_\odot/{\rm pc}^3$
with very high significance
(\cite{1995Natur.373..127M,1996Natur.383..415E,1998ApJ...508L..61L}).

Complementary observations of galactic nuclei with very long baseline
interferometry (VLBI) reveal the presence of compact radio cores
(\cite{1997ARA&A..35..607Z}) which appear to be coincident with the
central black hole candidates.  An intriguing case is that of the
Galactic Center where the bright, compact radio source Sgr A* lies at
the dynamical origin
(\cite{1997ApJ...475L.111M,1998ApJ...509..678G}). The nature of Sgr A*
is still unclear, since its structure is completely washed out by
strong interstellar scattering at cm-wavelengths (\cite{1998ApJ...508L..61L}). 
It is only at millimeter-wavelengths that we may begin
to see some internal structure
(\cite{1998ApJ...496L..97B,1998A&A...335L.106K,1998ApJ...508L..61L}).
Though the dark mass concentration could in principle be distributed
in the form of exotic objects on a scale slightly larger than the size
of Sgr A*---but with difficulties accounting for its radiation
characteristics (\cite{1999ApJ...511..750M})---it is expected to be
associated with Sgr A* itself, since the latter, unlike the
surrounding stars, has a tightly restricted proper motion indicating
that it is very heavy
(\cite{1998cpgc.workE...3R,1997MNRAS.291..219G}).

The key spectral features of Sgr A* are a slightly inverted
cm-wavelength spectrum, an apparent excess (or bump) at sub-millimeter
(sub-mm) wavelengths, and a steep cut-off towards the infrared
(\cite{1998ApJ...499..731F,1997ApJ...490L..77S}). The radio emission
is circularly polarized but undetected in linear polarization
(\cite{1999ApJ...521..582B,Bower1999c}). Proposed models for the radio
emission range from quasi-spherical inflows
(\cite{1992ApJ...387L..25M,1994ApJ...426..577M,1995Natur.374..623N})
to a jet-like outflow
(\cite{1993A&A...278L...1F,1999A&A...342...49F}).

The sub-mm bump is particularly interesting since this should be the
signature of a very compact synchrotron emitting region with a size of
a few Schwarzschild radii
(\cite{1996IAUS..169..169F,1998ApJ...499..731F}). The presence of
compact radio emission in Sgr A* at a wavelength as short as 1.4 mm
has been confirmed recently by a first VLBI detection at this
wavelength (\cite{1998A&A...335L.106K}). This detection is exciting
for several reasons.  First, it lies in a region of the spectrum where
the intrinsic source size should become apparent over
scatter-broadening by the intervening screen
(\cite{1992ApJ...395L..87M}).  Second, this component is sufficiently
bright to be detected with VLBI techniques at even shorter
wavelengths, and third, Sgr A* is sufficiently close that the size
scale where general relativistic effects are significant could be
resolved with VLBI at sub-mm wavelengths.  In addition, at sub-mm
wavelengths, the various models predict that the synchrotron emission
is not self-absorbed, allowing a view into the region near the
horizon. The horizon has a size of $(1+\sqrt{1-a_*^2})R_g$, where
$R_g\equiv GM/c^2$, $M$ is the mass of the black hole, $G$ is Newton's
constant, $c$ the speed of light, $a_*\equiv Jc/(GM^2)$ is the
dimensionless spin of the black hole in the range 0 to 1, and $J$ is
the angular momentum of the black hole.

Bardeen (1973)\nocite{Bardeen1973} described the idealized appearance
of a black hole in front of a planar emitting source, showing that it
literally would appear as a `black hole'. At that time such a
calculation was of mere theoretical interest and limited to just
calculating the envelope of the apparent black hole.  To test whether
there is a realistic chance of seeing this `black hole' in Sgr A*
(\cite{1998ApJ...499..731F}), we here report the first calculations
obtained with our general relativistic (GR) ray-tracing code that
allows us to simulate observed images of Sgr A* for various
combinations of black hole spin, inclination angle, and morphology of
the emission region directly surrounding the black hole and not just
for a background source. A more detailed description of our
calculations is in preparation (\cite{AFM99}).

\section{The appearance of a black hole}

We determine the appearance of the emitting region around a black hole
under the condition that it is optically thin.  For Sgr A* this might
be the case for the sub-mm bump (\cite{1998ApJ...499..731F}) indicated
by the turnover in the spectrum, and can always be achieved by going
to a suitably high frequency. Here we simply assume that the overall
specific intensity, $I_\nu$, observed at infinity is an integration of
the emissivity, $j_\nu$, times the differential path length along
geodesics (\cite{1997A&A...326..419J}).  In line with the qualitative
discussion of this paper, we assume that $j_\nu$ is independent of
frequency, and that it is either spatially uniform, or scales as
$r^{-2}$.  These two cases cover a large range of conditions expected
under several reasonable scenarios, be it a quasi-spherical infall, a
rotating thick disk, or the base of an outflow.


The calculation of the photon trajectories and the intensity
integrated along the line-of-sight is based on the standard formalism
(\cite{1981MNRAS.194..439T,1993A&A...272..355V,1994ApJ...421...46R,1997A&A...326..419J}).
Our calculations take into account all the well-known relativistic
effects, e.g., frame dragging, gravitational redshift, light bending,
and Doppler boosting.  The code is valid for all possible spins of the
black hole and for any arbitrary velocity field of the emission
region.

For a planar emitting source behind a black hole, a closed curve on
the sky plane divides a region where geodesics intersect the horizon
from a region whose geodesics miss the horizon (\cite{Bardeen1973}).
This curve, which we refer to as the ``apparent boundary'' of the
black hole, is a circle of radius $\sqrt{27} R_g$ in the Schwarzschild
case ($a_*=0$), but has a more flattened shape of similar size for a
Kerr black hole, slightly dependent on inclination.  The size of the
apparent boundary is much larger than the event horizon due to strong
bending of light by the black hole.  When the emission occurs in an
optically thin region {\em surrounding} the black hole, the case of
interest here, the apparent boundary has the same exact shape since
the properties of the geodesics are independent of where the sources
are located.  However, photons on geodesics located within the
apparent boundary that can still escape to the observer experience
strong gravitational redshift and a shorter total path length, leading
to a smaller integrated emissivity, while photons just outside the
apparent boundary can orbit the black hole near the circular photon
radius several times, adding to the observed intensity
(\cite{1997A&A...326..419J}).  This produces a marked deficit of the
observed intensity inside the apparent boundary, which we refer to as
the ``shadow'' of the black hole.

We here consider a compact, optically-thin emitting region surrounding a 
black hole with spin parameter $a_*=0$ (i.e., a Schwarzschild black hole) 
and a maximally spinning Kerr hole with $a_*=0.998$.  In the set of
simulations shown here, we take the viewing angle $i$ to be $45^\circ$
with respect to the spin axis (when it is present), and we consider
two distributions of gas velocity $v$. The first has the plasma in
free-fall, i.e., $v^r=-\sqrt{2r(a^2+r^2)}\Delta/A$ and $\Omega =
2ar/A$, where $v^r$ is the Boyer-Lindquist radial velocity, 
$\Omega$ is the orbital frequency, $\Delta\equiv r^2-2r+a^2$, and
$A\equiv(r^2+a^2)^2-a^2\Delta\sin^2{\theta}$. (We have set $G=M=c=1$
in this paragraph.)  The second has the plasma orbiting in rigidly
rotating shells with the equatorial Keplerian frequency $\Omega =
1/(r^{3/2}+a)$ for $r>r_{ms}$ with $v^r=0$, and infalling with
constant angular momentum inside $r<r_{ms}$ (\cite{1975ApJ...202..788C}),
with $v^\theta=0$ for all $r$.  


In order to display concrete examples of how realistic our proposed
measurements of these effects with VLBI will be, we have simulated the
expected images for the massive black hole candidate Sgr A* at the
Galactic Center.  For its measured
mass (\cite{1996Natur.383..415E,1998ApJ...509..678G}) $M = 2.6\times
10^6\;M_\odot$, the scale size for this object is the gravitational
radius $R_g=3.9\times 10^{11}$ cm, which is half of the Schwarzschild
radius $R_s\equiv 2GM/c^2$.

To simulate an observed image we have to take two additional effects
into account: interstellar scattering and the finite telescope
resolution achievable from the ground. Scatter-broadening at the
Galactic Center is incorporated by smoothing the image with an
elliptical Gaussian with a FWHM of 24.2
$\mu$arcsecond$\times(\lambda/1.3\,\mbox{mm})^{2}$ along the major axis
and 12.8 $\mu$arcsecond$\times(\lambda/1.3\,\mbox{mm})^{2}$ along the
minor axis (\cite{1998ApJ...508L..61L}). The position angle of this
ellipse is arbitrary since we do not know yet the spin axis of the
black hole on the sky and we have assumed PA=$90^\circ$ for the major
axis. The telescope resolution---in an idealized form---is then added by
convolving the smoothed image with a spherical Gaussian point-spread
function of FWHM 33.5 $\mu$arcsecond$\times(\lambda/1.3\,\mbox{mm})^{-1}
(l/8000\mbox{km})^{-1}$---the possible resolution of a global
interferometer with 8000 km baselines (\cite{Kri1996}). In reality the
exact point-spread-function will of course depend on the number and
placement of the participating telescopes.

\epsscale{0.85}
\begin{figure*}[htb]
\figurenum{1}
\psfig{figure=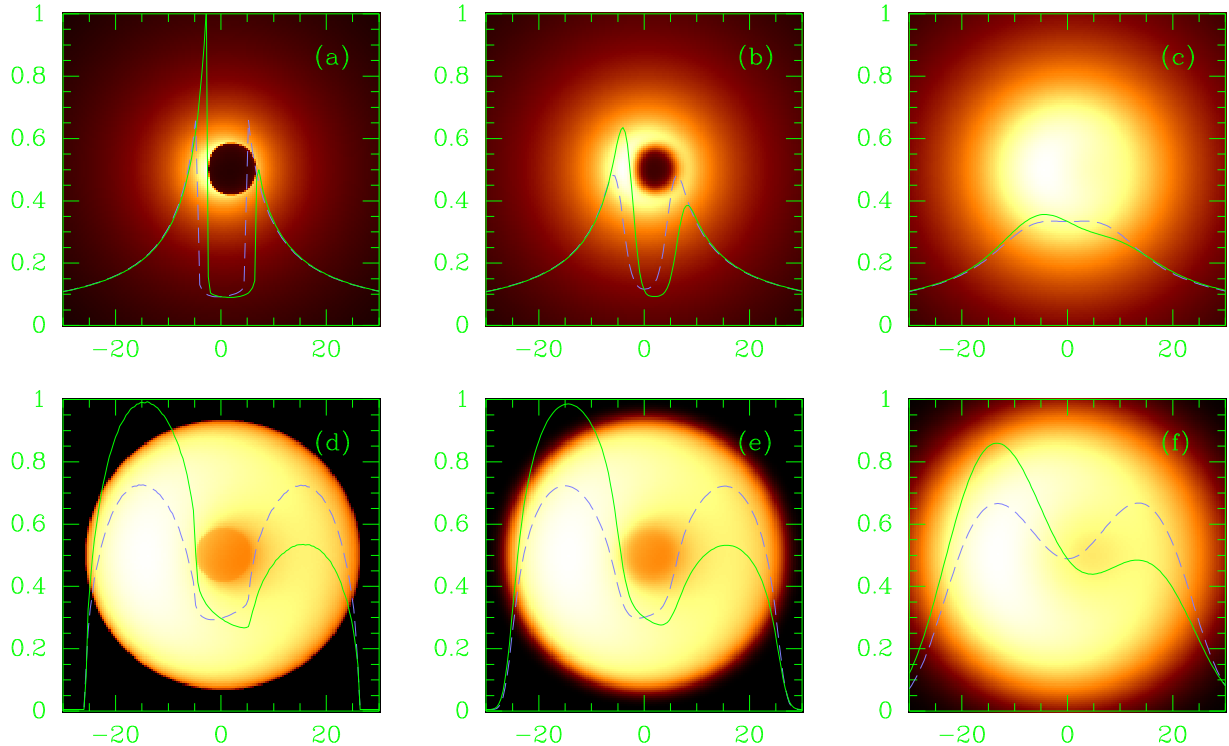,width=0.9\textwidth,bblly=11.2cm,bbury=21.5cm,bbllx=0.8cm,bburx=18.2cm}
\caption{
An image of an optically thin emission region surrounding a black hole
with the characteristics of Sgr A* at the Galactic Center.  The black
hole is here either maximally rotating ($a_* = 0.998 $, Figs.~1a-c) or
non-rotating ($a_*=0$, Figs.~1d-f). The emitting gas is assumed to be
in free fall with an emissivity $\propto r^{-2}$ (top) or on Keplerian
shells (bottom) with a uniform emissivity (viewing angle
$i=45^\circ$). Figs.~1a\&d show the GR ray-tracing calculations,
Figs.~1b\&e are the images seen by an idealized VLBI array at
0.6 mm wavelength taking interstellar scattering into account, and
Figs.~1c\&f are those for a wavelength of 1.3 mm. The intensity
variations along the $x$-axis (solid green curve) and the $y$-axis
(dashed purple curve) are overlayed. The vertical axes show the
intensity of the curves in arbitrary units and the horizontal axes
shows the distance from the black hole in units of $R_{\rm g}$ which
for Sgr~A* is $3.9\times 10^{11}$ cm $\sim3\;\mu$arcseconds.}
\end{figure*}

In Figure 1, we show the resulting image of Sgr A* for a maximally
rotating black hole viewed at an angle of $i=45^\circ$, for a compact
region in free fall, with an emissivity $j_\nu= \nu^0r^{-2}$.  We
first show the original, unsmoothed image of the emission region as
calculated with the GR code in panel (a), and then present the
simulated `observed' images at 0.6 and 1.3 mm wavelengths in panels (b)
and (c), respectively. The two distinct features that are evident in
Figure 1a are (1) the clear depression in $I_\nu$---the
shadow---produced near the black hole, which in this particular
example represents a modulation of up to 90\% in intensity from
peak to trough, and (2) the size of the shadow, which here is
$9.2R_{\rm g}$ in diameter.  This represents a projected size of 27
$\mu$arcseconds, which is already within a factor of two of the
current VLBI resolution (\cite{Kri1995}).  The shadow is a generic
feature of various other models we have looked at, including those
with outflows, cylindrical emissivity, and various inclinations or
spins.

To illustrate the expected image for another extreme case, we show in
Figure 1d the analogue to Figure 1a for the case with $a_*=0$ (i.e.,
no rotation), an emitting plasma orbiting in Keplerian shells (as
described above), and a uniform $j_\nu$ for $r < 25 R_g$.  Even though
these conditions are distinctly different compared to those of Figure
1a, the black hole shadow is still clearly evident, here representing
a modulation in $I_\nu$ in the range of 50-75\% from peak to trough
(Fig.~1d), and with a diameter of roughly $10.4\,R_g$.  In this case,
the emission is asymmetric due to the strong Doppler shifts associated
with the emission by a rapidly moving plasma along the line-of-sight
(with velocity $v_\phi$).  


The important conclusion is that the diameter of the shadow---in
marked contrast to the event horizon---is fairly independent of the
black hole spin and is always of order 10$R_{\rm g}$.  Indeed, this is
consistent with the observed 0.8 mm size limit $>4 R_g$ of Sgr A* from
a lack of scintillation (\cite{1991ApJ...381L..43G}).  The presence of
a rotating hole viewed edge-on will lead to a shifting of the apparent
boundary (by as much as 2.5 $R_g$, or 8 $\mu$arcseconds) with respect
to the center of mass, or the centroid of the outer emission region.

Interestingly, the scattering size of Sgr A* and the resolution of
global VLBI arrays become comparable to the size of the shadow at
a wavelength of about 1.3 mm. As one can see from Figures 1c\&f the
shadow is still almost completely washed out for VLBI observations at
1.3 mm, while it is very apparent at a factor two shorter wavelength
(Figures 1b\&e). In fact, already at 0.8 mm (not shown here) the shadow
can be easily seen. Under certain conditions, i.e., a very homogeneous
emission region, the shadow would be visible even at 1.3 mm (Fig.~1f).

\section{How realistic is such an experiment?}

The arguments for the feasibility of such an experiment are rather
compelling. First of all, the mass of Sgr A* is very well known
within 20\%, the main uncertainty being the exact distance to
the Galactic Center. Since, as we have shown, the unknown spin of the
suspected black hole contributes only another 10\% uncertainty, we can
conservatively predict the angular diameter of the shadow in Sgr A*
from the GR calculations alone to be $\sim30\pm7 \mu$arcseconds
independent of wavelength. As seen in Fig.~1, the finite telescope
resolution and the scatter broadening will make the detectability of
the shadow a function of wavelength and emissivity; however, the size
of the shadow will remain of similar order and under no circumstances
can become smaller.

The technical methods to achieve such a resolution at wavelengths
shortwards of 1.3 mm are currently being developed and a first
detection of Sgr A* at 1.4 mm with VLBI has already been reported. The
challenge will be to push this technology even further towards 0.8 or
even 0.6 mm VLBI. Over the next decade many more telescopes are
expected to operate at these wavelengths.  Depending on how short a
wavelength is required, the projected time scale for developing the
necessary VLBI techniques may be about ten years.  A fundamental
problem preventing such an experiment is not now apparent, but in
light of our results, planning of the new sub-mm-telescopes should
include sufficient provisions for VLBI experiments.

A potential problem with our model may occur if $j_\nu$ has an inner
cutoff which is larger than that of the horizon, making the shadow
larger than predicted due to a decrease in emissivity rather than to
GR effects.  However, first of all, the truncation
of accretion disk emission at the marginal stable orbit $r_{\rm ms}$
is somewhat arbitrary (\cite{1975ApJ...202..788C}) and, secondly, if
it exists such a cutoff would likely be frequency dependent, while
there will be a frequency-independent minimum radius due to the
general relativistic effects we have described.  Another problem could
be the unknown morphology of the emission region. Anisotropy, strong
velocity fields, and density inhomogeneities would make an
identification of the shadow in an observed image more
difficult. However, inhomogeneities are unlikely to be a major issue,
since the time scale for rotation around the black hole in the
Galactic Center is only a few hundred seconds and hence much less than
the typical duration of a VLBI observation. The strong shear near the
black hole would tend to smooth out any inhomogeneities very
quickly. Indeed, sub-mm variability studies on such short time scales
(\cite{1991ApJ...381L..43G}) have yielded negative results. The same
argument applies to emission models which are offset from the black
hole, e.g., are one-sided. Since the shadow of the black hole has a
very well defined shape it would under any conditions appear as a
distinct feature, given that the dynamic range of the map is large
enough (i.e., $\ga$100:1, considering a range of emission models,
\cite{AFM99}).

Finally, synchrotron self-absorption could pose a problem. So far the
available sub-mm spectra show a flattening of the spectrum around
1.3-0.6 mm indicating a turnover towards an optically thin spectrum.
Given the current observational uncertainties one could in principle
construct simple models where the flow does not become optically thin
until 0.2 mm.  Improved simultaneous measurements at sub-mm wavelengths
are therefore highly desirable to exactly measure the spectral
turnover since the experiment we propose here will only work for an
optically thin flow.  At hundreds of microns the atmosphere becomes
optically thick, making much more expensive space-based observations
necessary.  At X-ray wavelengths, the accretion flow will be optically
thin to electron scattering, so there may be a better chance of
detecting the shadow with future space-based X-ray interferometry as
proposed in the MAXIM experiment.

\section{Summary}
The importance of the proposed imaging of Sgr A* at sub-mm wavelengths
with VLBI cannot be overemphasized.  The bump in the spectrum of Sgr
A* strongly suggests the presence of a compact component whose
proximity to the event horizon is predicted to result in a shadow of
measurable dimensions in the intensity map.  To our knowledge, such a
feature is unique and Sgr~A* seems to have all the right parameters to
make it observable.  The observation of this shadow would confirm the
widely held belief that most of the dark mass concentration in the
nuclei of galaxies such as ours is contained within a black hole, and
it would be the first direct evidence of the existence of an event
horizon. A non-detection with sufficiently developed techniques, on
the other hand, might pose a major problem for the standard black hole
paradigm. Because of this fundamental importance, the
experiment we propose here should be a major motivation for
intensifying the current development of sub-mm astronomy in general
and mm- and sub-mm VLBI in particular.

{\bf Acknowledgments} We thank P.L. Biermann. T. Krichbaum,
A. Zensus, O. Blaes, R. Antonucci, and M. Reid for useful
discussions. This work was supported in part by a Sir Thomas Lyle
Fellowship (FM), NASA grant NAG58239 (FM), DFG grants Fa 358/1-1\&2
(HF), and NSF grant AST-9616922 (EA).  EA would like to thank the ITP
at the University of California at Santa Barbara for their
hospitality.



\end{document}